\documentclass[12pt,twoside]{article}   %For printing on two sides of page
\usepackage[super,sort&compress,comma]{natbib}
\usepackage{fancyhdr}		%Gives headers and footers defined below
\usepackage[section]{placeins}   %
\usepackage{graphicx}
\graphicspath{{./figures/}}

\makeatletter \renewcommand\@biblabel[1]{$^{#1}$} \makeatother
\newcommand{\cen}[1]{\begin{center} #1 \end{center}}

\usepackage[mathlines]{lineno}
\usepackage{hyperref}
\hypersetup{ colorlinks,
    citecolor=blue,
    filecolor=blue,
    linkcolor=blue,
    urlcolor=blue
}

\usepackage{xcolor}
\definecolor{gray}{rgb}{0.6,0.6,0.6}
\definecolor{red}{rgb}{0.85,0,0}
\definecolor{green}{rgb}{0,0.85,0}
\definecolor{blue}{rgb}{0,0,0.85}
\definecolor{beige}{rgb}{0.92,0.87,0.78}
\usepackage[all]{hypcap}    %causes link to figures to go to figure, not caption
\usepackage{setspace}
\usepackage{amssymb}
\usepackage{amsmath}
\usepackage{multirow}
\usepackage{booktabs}
\usepackage{enumerate}
\usepackage{authblk}
\usepackage{lineno}
\usepackage{stfloats}

\begin{document}

\cen{\sf {\Large {\bfseries Multi-Energy Blended CBCT Spectral Imaging Using a Spectral Modulator with Flying Focal Spot (SMFFS) } \\  
\vspace*{10mm}
Yifan~Deng$^{1,2}$, Hao Zhou$^{1,2}$, Zhilei Wang$^{1,2}$, Adam S. Wang$^{3}$, and Hewei~Gao$^{*1,2}$} \\
$^{1}$Department of Engineering Physics, Tsinghua University, Beijing 100084, China\\
$^{2}$Key Laboratory of Particle \& Radiation Imaging (Tsinghua University), Ministry of Education, China\\
$^{3}$Department of Radiology, Stanford University, Stanford, CA 94305, USA\\
%\vspace{5mm}\\
%Version typeset \today
}

\pagenumbering{roman}
\setcounter{page}{1}
\pagestyle{plain}
Author to whom correspondence should be addressed. email: hwgao@tsinghua.edu.cn \\
% note, probably best not to use a student's e-mail as it won't be valid for
% very long.

\begin{abstract}
\noindent 
{\bf Background:} Cone-beam CT (CBCT) has been extensively employed in industrial and medical applications, such as image-guided radiotherapy and diagnostic imaging, with growing demand for quantitative imaging using CBCT. However, conventional CBCT can be easily compromised by scatter and beam hardening artifacts, and the entanglement of scatter and spectral effects introduces additional complexity.\\
{\bf Purpose:} 
The intertwined scatter and spectral effects within CBCT pose significant challenges to the quantitative performance of spectral imaging. In this work, we present the first attempt to develop a stationary spectral modulator with flying focal spot (SMFFS) technology as a promising, low-cost approach to accurately solving the X-ray scattering problem and physically enabling spectral imaging in a unified framework, and with no significant misalignment in data sampling of spectral projections.\\ 
{\bf Methods:} 
To deal with the intertwined scatter-spectral challenge, we propose a novel scatter-decoupled material decomposition (SDMD) method for SMFFS, which consists of four steps in total, including 
1) spatial resolution-preserved and noise-suppressed multi-energy ``residual'' projection generation free from scatter, based on a hypothesis of scatter similarity; 
2) first-pass material decomposition from the generated multi-energy residual projections in non-penumbra regions, with a structure similarity constraint to overcome the increased noise and penumbra effect; 
3) scatter estimation for complete data; and
4) second-pass material decomposition for complete data by using a multi-material spectral correction method.
Monte Carlo simulations of a pure-water cylinder phantom with different focal spot deflections are conducted to validate the scatter similarity hypothesis. 
Both numerical simulations using a clinical abdominal CT dataset, and physics experiments on a tabletop CBCT system using a Gammex multi-energy CT phantom and an anthropomorphic chest phantom, are carried out to demonstrate the feasibility of CBCT spectral imaging with SMFFS and our proposed SDMD method. \\
{\bf Results:} 
Monte Carlo simulations show that focal spot deflections within a range of ~2 mm share quite similar scatter distributions overall. Numerical simulations demonstrate that SMFFS with SDMD method can achieve better material decomposition and CT number accuracy with less artifacts. 
In physical experiments, for the Gammex phantom, the root mean square error (RMSE) in selected regions of interest (ROIs) of virtual monochromatic image (VMI) at 70 keV is 19.2 HU in SMFFS cone-beam(CB) scan, and 24.1 and 214.3 HU in sequential 80/120 kVp (dual kVp, DKV) CB scan with and without scatter correction, respectively. For the chest phantom, the RMSE in selected ROIs of VMIs is 11.8 HU for SMFFS CB scan, and 14.5 and 437.6 HU for sequential 80/140 kVp (DKV) CB scan with and without scatter correction, respectively. Also, the non-uniformity among selected regions of the chest phantom is 14.1 HU for SMFFS CB scan, and 59.4 and 184.0 HU for the DKV CB scan with and without a traditional scatter correction method, respectively. 
\\
{\bf Conclusions:} 
We propose a scatter-decoupled material decomposition (SDMD) method for CBCT with SMFFS. Our preliminary results show that SMFFS can enable spectral imaging with simultaneous scatter correction for CBCT and effectively improve its quantitative imaging performance. \\

\end{abstract}
%\note{This is a sample note.}

\newpage     %may or may not be needed

%The table of contents is for drafting and refereeing purposes only. Note that all links to references, tables and figures can be clicked on and returned to calling point using cmd[ on a Mac using Preview or some equivalent on PCs (see View - go to on whatever reader).
\tableofcontents

\newpage

\setlength{\baselineskip}{0.7cm}      %double spacing		

\pagenumbering{arabic}
\setcounter{page}{1}
\pagestyle{fancy}

\section{Introduction}
Cone-beam computed tomography (CT) utilizes a large-area flat-panel detector (FPD) that can provide better system integration, higher spatial resolution, lower equipment cost, and more flexibility.
In recent years, cone-beam CT (CBCT) imaging has been widely investigated and applied into medical and industrial applications, such as image-guided radiotherapy (IGRT)\cite{Posiewnik2019IGRT}, dental CT\cite{Dental2021review}, breast CT\cite{Connell2021CBCTBreast}, etc, with increasing demand of improvement in image quality. However, due to the physics limitations, conventional CBCT can be easily compromised by scatter, beam hardening, and cone-beam artifacts, etc. 

Spectral CT technology is able to achieve much better quantitative performance as it allows material decomposition by taking two or more measurements under different effective X-ray spectra, enabled by using a  source-based, detector-based, or filter-based dual- or multi-energy spectral approach. Dual-energy CT equipped with the dual-source, fast-kV switching, dual-layer detector, and splitting-filter technologies have already been successfully applied into clinical applications by leading multi-detector  CT vendors\cite{McCollough2020multi}; 
Multi-energy CT could be achieved by using a photon-counting detector (PCD) that has advanced greatly nowadays, with the very first clinical scanner being approved recently \cite{Rajendran2022FirstClinicalPCD}. 

For cone-beam CT, spectral imaging is also highly desired and is under active investigation with great progress these days. Numerous spectral CBCT concepts and prototype systems have been developed, which can mainly be grouped into the source-based, such as the fast kV-switching\cite{Muller2016kVCBCT}\cite{Cassetta2020kVCBCT}, and the detector-based, such as the dual-layer detector\cite{ShiDLCBCT2020}\cite{Stahl2021DLCBCT} and photon counting detector (PCD)\cite{Ji2021PCDCBCT} . 
These investigations have shown the potential of spectral CBCT in diagnostic and interventional imaging. However, as one of the most important factors affecting spectral CBCT performance, scatter was either simply estimated and corrected or completely avoided by using a relatively narrowed collimator, in most of the studies in the literature.
Unfortunately the performance of anti-scatter grid for FPD is usually compromised in suppressing the scattered signals while preserving the primary signals. As a result, the integration of spectral technology into the CBCT system to achieve better quantitative performance in practical applications, still remains a sophisticated task, with many challenges to overcome. 

Using X-ray source modulation methods could also be a promising way of realizing quantitative CBCT spectral imaging. In the literature, several modulation-based methods for spectral imaging have also been explored, such as the splitting filter for twin-beam dual-energy CT proposed in 2016\cite{Euler2016Inv}, the primary modulator for single-scan dual-energy CT proposed in 2018\cite{Petrongolo2018}, 
the spatial-spectral modulator for material decomposition proposed in 2019\cite{Tivnan2019SPIE}, the spectral modulator with flying focal spot for spectral CT proposed in 2019\cite{Gao2019SMFFS}, the anti-scatter grid as a filter with flying focal spot for dual-energy CT in 2021\cite{Hsieh2021FFS}, and the fine grid structure with a varying X-ray incidence angle for spectral CT in 2021\cite{Stayman2021MP}. These methods intended to use a low-cost splitting filter or modulator to generate different spectra in the spatial domain and eventually achieve promising performance for spectral imaging. Moreover, modulation-based methods have great potential in scatter correction. In 2006, using the primary modulator for scatter correction in the frequency domain first emerged\cite{zhu2006scatter, Gao2010scatter}. Subsequently, more robust methods based on the modulator have also been developed\cite{Ritschl2015}\cite{Chen2016modu} and achieved better performance in photon-counting detector CT\cite{PivotTMI2020} and dual-layer detector\cite{shi2022singleshot}. 
However, in these studies, the scatter correction and spectral imaging were separately handled.

In order to promote CBCT quantitative imaging, our group has proposed a novel concept of the stationary spectral modulator with flying focal spot (SMFFS)\cite{Gao2019SMFFS,Deng2020CTMeeting}, which is low-cost, enables scatter correction and spectral imaging\cite{Deng2023fully3d}, and is flexible to be integrated into other existing spectral CT systems.
Flying focal spot (FFS) is a technique that allows view-by-view deflections of the focal spot in the x- and/or z-direction, utilized to reduce aliasing artifacts and increase data sampling rate \cite{KachelrieTMI2006}.
Compared with other modulator-based approaches\cite{Petrongolo2018,TivnanMP2021}, SMFFS maintains a stationary filter for easier electromechanical control, and can generate multi-energy blended data with a similarity in scatter distribution and no significant misalignment of spectral projections thanks to the use of FFS\cite{GaoSPIE2020}. In the previous work, we investigated its feasibility of spectral correction and scatter correction, using a copper modulator\cite{GaoMP2021}.
%\cite{GaoMP2021,ZhangPMB2021}.

In this work, we further advance this SMFFS technology to spectral imaging with a mixed two-dimensional (2D) molybdenum modulator, 
and develop a novel scatter-decoupled material decomposition method, where scatter correction and spectral imaging are modeled in a unified framework, and a strong similarity of scatter distributions in SMFFS is taken advantage of. Both numerical simulations and physical experiments validate the effectiveness of our proposed method.

\section{Multi-Energy Blended CBCT Spectral Imaging}
\subsection{Stationary Spectral Modulator with Flying Focal Spot}
\begin{figure}[htb]
	\centering
	\includegraphics[width=70mm]{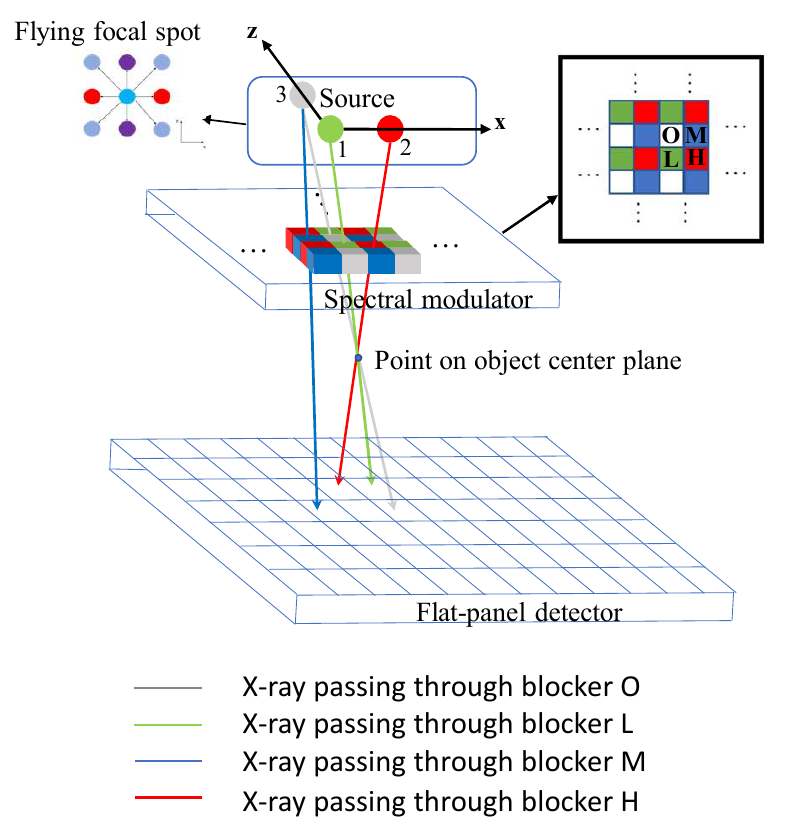}
	\caption{An illustrative diagram of the spectral modulator with flying focal spot (SMFFS), where multi-energy X-rays can be generated by the modulator spatially, and misalignment of multi-energy X-rays can be significantly reduced by flying focal spot.} \label{fig:SMFFS_system}
\end{figure}

Figure~\ref{fig:SMFFS_system} shows an illustration of an SMFFS system, which mainly consists of the X-ray source, the spectral modulator, the scanning object, and the flat-panel detector. In such a system, 
the X-ray source should be able to equivalently deflect the focal spot during a CT scan using the FFS technology (or alternatively, with a distributed X-ray source); 
the detector can be an energy-integrating flat-panel detector in this work;
the spectral modulator consists of partially attenuating blockers and is placed between the X-ray source and the scanning object. 

The SMFFS system has the potential for multi-energy spectral imaging with scatter correction capability. For spectral imaging, a spectral modulator does not entirely block the intensities of the X-ray beams, while taking advantage of the energy-dependent attenuation property of the modulator for polychromatic X-rays, and thanks to the focal spot deflection, the filter pattern of the stationary spectral modulator can change from one view to another during a CT scan with easier electromechanical control and good alignment for multi-energy X-rays. For scatter correction, due to the low-frequency property of the scatter distribution and the slight wobbling of the focal spot (about 1 mm), 
the scatter distributions can be kept similar across different focal spot positions. 

For the SMFFS system with the focal spot at the $k$-th position, in consideration of the scatter intensity $I_{s}^{(k)}$ received on the detector, the total X-ray intensity measurement with an object in the beam can be written as,
\begin{equation}
	\begin{aligned}
		I_{t}^{(k)} &= \int S^{(k)}(E)e^{-\mu_1(E)L_{1}^{(k)}-\mu_2(E)L_{2}^{(k)}} \text{d}E+ I_{s} ^{(k)}
		\label{eq:It_model0}
	\end{aligned}
\end{equation}
Here the superscript, $(k)$, corresponds to the different focal spot positions that can greatly avoid the misalignment of spectral projections as shown in Fig. \ref{fig:SMFFS_system}; 
$S^{(k)}(E)$ is the effective spectra passing through the modulator and received by the detector (without object); 
$\mu_1(E)$ and $\mu_2(E)$ denote the energy-dependent attenuation coefficients of two basis materials, and $L_1^{(k)}$ and $L_2^{(k)}$ represent the corresponding effective path-lengths of two basis materials, respectively. 
$S^{(k)}(E)$ can be further expressed as,
\begin{equation}
	S^{(k)}(E) = S_O(E)e^{-\sum_i \mu_{m_i}(E) L_{m_i}^{(k)}} D(E)
\end{equation}
where, $S_O(E)$ is the original spectrum without filtration of the modulator; $\mu_{m_i}$ is the linear attenuation coefficient of the material $i$ of the modulator; $L_{m_i}^{(k)}$ is the corresponding path-length of the material $i$ at the $k$-th focal spot position; $D(E)$ is the energy response of the detector.

%-------------------------------------------------------------------------------------------
\subsection{Potential of Spectral CBCT Imaging with Scatter Similarity Hypothesis}
Given the fact that scattered signals are mostly low-frequency dominated, and the deflection of the focal spot is only about 1 mm in SMFFS, it is reasonable to assume that before and after wobbling the focal spot, the scatter distributions will be very similar ($I_{s}\approx I_{s}^{(k)}$). Also, the SMFFS system can maintain a good alignment of multi-energy X-rays among selected focal spot positions. Thus, the equivalent path-lengths of basis materials $L_{1}^{(k)}, L_{2}^{(k)}$ and the scatter intensities $I_{s}^{(k)}$ at different focal spot positions can be approximately considered as matched. Therefore, the total X-ray intensity measurements in Eq. \eqref{eq:It_model0} can be simplified as, 
\begin{equation}
	\begin{aligned}
		I_{t}^{(k)} &= \int S^{(k)}(E)e^{-\mu_1(E)L_{1}-\mu_2(E)L_{2}} \text{d}E + I_{s}
		\label{eq:It_model}
	\end{aligned}
\end{equation}
It is seen that there are only three unknowns in Eq. \eqref{eq:It_model}: $L_1, L_2, I_s$.

\begin{figure}[tb]
	\centering
	\includegraphics[width=65mm]{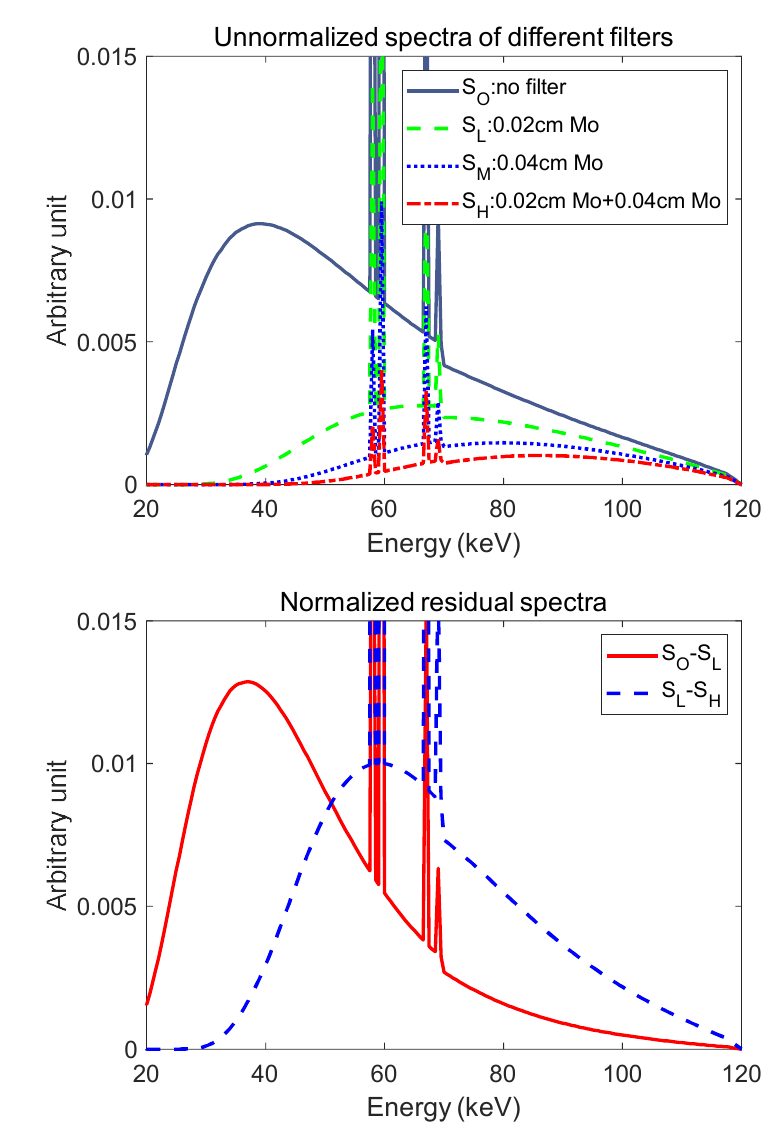}
	\caption{Simulated effective spectra of an X-ray beam passing through different blockers of a spectral modulator with ``Mo+Mo'' combination. Top: effective spectra after filtration of different blockers; Bottom: residual spectra after subtractions.} \label{fig:spectrum_plot}
\end{figure}

If multiple measurements at different effective spectra are available, Eq. \eqref{eq:It_model} can be mathematically solved in a unified framework. A straightforward but perhaps non-optimal approach can be like this,
\begin{equation}
	\begin{aligned}
		I_{t}^{(i)} - I_{t}^{(k)} = \int \left(S^{(i)}(E)-S^{(k)}(E)\right) e^{-\sum_i \mu_i(E)L_{i}} \text{d}E 
		\label{sub_model}
	\end{aligned}
\end{equation}
where $S^{(i)}(E)-S^{(k)}(E)$ is regarded as the residual spectrum, and $I_{t}^{(i)} - I_{t}^{(k)}$ the residual primary data in SMFFS.
Fig.~\ref{fig:spectrum_plot} shows a simulated residual spectra and the filtered spectra by the 2D modulator used in this work with an experimental calibrated 120 kVp spectrum used as the initial spectrum. It is seen that the residual spectra can maintain high spectral diversity thanks to the spectra generated by filtration of the specific blockers. 

It should be noted that for a specific detector pixel $idx_1$ under focal spot position $1$, and its corresponding pixel $idx_2$ under focal spot position $2$ with x-direction deflection or $idx_3$ under focal spot position $3$ with z-direction deflection, their effective spectra are generated by different filtrations of the specific blockers as shown in Fig. \ref{fig:SMFFS_system}. For example, in this work, we use $S_O, S_L, S_M, S_H$ for the spectrum of X-ray passing through the blocker $O, L, M, H$, respectively. Specifically, one can obtain different spectra at adjacent detector pixels under different focal spot position. For example, for x-FFS to deflect focal spot from position 1 to position 2, 
\begin{equation*}
	\begin{aligned}
		S^{(1)}_{idx_1}(E)&=S_O(E), S^{(2)}_{idx_2}(E) =S_L(E), \\
		\text{or} \; S^{(1)}_{idx_1}(E)&=S_M(E), S^{(2)}_{idx_2}(E)=S_H(E) \\
	\end{aligned}
\end{equation*}
and for z-FFS to deflect focal spot from position 1 to position 3, 
\begin{equation*}
	\begin{aligned}
		S^{(1)}_{idx_1}(E)&=S_O(E), S^{(3)}_{idx_3}(E)=S_M(E), \\
		\text{or} \; S^{(1)}_{idx_1}(E)&=S_L(E), S^{(3)}_{idx_3}(E)=S_H(E)\\
	\end{aligned}
\end{equation*}
Other combinations can also be produced according to the modulator patterns. 

\section{Scatter-Decoupled Material Decomposition}
\subsection{Scatter-Spectral Decoupling in Projection Domain}
\label{subsec:model with scatter}
As Eq. \eqref{sub_model} shows, $I_{t}^{(i)} - I_{t}^{(k)}$ becomes residual data free from scatter, and the specific modulator pattern with a specific FFS distance can make the spectral projections from different focal spot positions passing through different blockers with good alignment as Fig. \ref{fig:SMFFS_system} shows. Thus, we first divide the initial sets of data into different energy levels corresponding to different filtrations. Taking the detector pixels with $I_{t}^{(1)} = I_{t_O},  I_{t}^{(2)} =  I_{t_L}, I_{t}^{(3)} = I_{t_H}$ as an example, we can obtain a pair of residual projections without scatter, $P_{OL}, P_{LH}$, as,
\begin{equation}
	\begin{aligned}
		P_{OL} &= -\ln{\left(\frac{I_{t}^{(1)} - I_{t}^{(2)}}{I_{m}^{(1)} - I_{m}^{(2)}}\right)}= -\ln{\left(\frac{I_{t_O} - I_{t_L}}{I_{m_O} - I_{m_L}}\right)}\\
		&=-\ln{\left(\frac{\int\Delta{S_{OL}(E)}e^{-\sum_{i=1}^2 \mu_i(E)L_i} \text{d}E }{\int\Delta{S_{OL}(E)}\text{d}E }\right)}\\
		P_{LH} &= -\ln{\left(\frac{I_{t}^{(2)} - I_{t}^{(3)}}{I_{m}^{(2)} - I_{m}^{(3)}}\right)} =-\ln{\left(\frac{I_{t_L} - I_{t_H}}{I_{m_L} - I_{m_H}}\right)}
		\\ &=-\ln{\left(\frac{\int\Delta{S_{LH}(E)}e^{-\sum_{i=1}^2 \mu_i(E)L_i} \text{d}E }{\int\Delta{S_{LH}(E)}\text{d}E }\right)}
		\label{sub_model_all}
	\end{aligned}
\end{equation}
where, $I_{m}$ is the X-ray intensity measured with the modulator but without the scanning object in the beam; the subscript of $I_m$ and $I_t$ represents the energy level (O: original spectrum without passing through the modulator, L: low filtration by the modulator, M: middle filtration, H: high filtration), corresponding to different blockers of the modulator.

Unfortunately, the signal noise in the residual projection $P_{OL},P_{LH}$ will be significantly increased by the subtraction, 
the low signals of high-energy data,
and the noise of scattered signals.
In order to suppress noise, Eq. \eqref{sub_model_all} can be split into the initial projection term and scatter suppressing terms as,
\begin{equation}
	\begin{aligned}
		P_{OL} &= P_O-\ln{\left(\frac{I_{m_O}}{I_{m_O}-I_{m_L}}\right)}
		-\ln{\left(1-\frac{I_{t_L}}{I_{t_O}}\right)} \\
		P_{LH} &=  P_L-\ln{\left(\frac{I_{m_L}}{I_{m_L}-I_{m_H}}\right)}
		-\ln{\left(1-\frac{I_{t_H}}{I_{t_L}}\right)} 
		\label{sub_model_split}
	\end{aligned}
\end{equation}
where, $P_O = -\ln{\left(\frac{I_{t_O}}{I_{m_O}}\right)}$ and $P_L=-\ln{\left(\frac{I_{t_L}}{I_{m_L}}\right)}$ are the initial projection with scatter. Taking $P_{OL}$ for example, $-\ln{\left(\frac{I_{m_O}}{I_{m_O}-I_{m_L}}\right)}$ can be regarded as a constant because it is independent of the scanning object. Therefore, $-\ln{\left(1-\frac{I_{t_L}}{I_{t_O}}\right)}$ is the key term for scatter suppression, and also the main factor of noise increase for residual data. In this paper, we use the Taylor Expansion of $-\ln{\left(1-\frac{I_{t_L}}{I_{t_O}}\right)}$ in Eq. \eqref{sub_model_split} and truncate it to the $N$-th term as $\sum_{n=1}^{N} \left(\frac{I_{t_L}}{I_{t_O}}\right)^n$. Then, we can low-pass filter the initial scatter suppressing term $-\ln{\left(1-\frac{I_{t_L}}{I_{t_O}}\right)}$ or the truncated terms $\left(\frac{I_{t_L}}{I_{t_O}}\right)^n$ to control the noise level and maintain the performance of scatter suppression, without degrading spatial resolution significantly.
In this paper, we use a gaussian filter with a size of $[5,5]$ pixels. Of course, the smoothing parameters need to be tuned in practical applications.

\subsection{Two-pass Guided Material Decomposition}
\label{sec:MD method}
After the noise reduction tactics used in the scatter-decoupling step, the material decomposition is still ill-posed due to the interpolation challenge for the residual data in the penumbra area. Therefore, we propose a two-pass guided material decomposition approach to solving this problem while preserving spatial resolution.

\subsubsection{First-pass Guided Material Decomposition for Residual Data} \label{subsubsec:firstpass}
Among the different combinations of residual projections, a particular combination is $P_{OH}= -\ln{\left(\frac{I_{t_O} - I_{t_H}}{I_{m_O} - I_{m_H}}\right)}$, which has the lowest noise level, but not the best choice for energy separation. Therefore, reconstruction from $P_{OH}$ can be used as a guided image for material decomposition using $P_{OL}, P_{LH}$. We conduct a guided material decomposition by referring to an iterative similarity-based method\cite{Wang2016PMB}. Then, we obtain the over-smooth basis material images $x_1, x_2$ by the residual data.

\subsubsection{Scatter Estimation} 
Taking the decomposition of iodine and water as an example, we use the first-pass, over-smooth iodine result $x_{io}$ from \ref{subsubsec:firstpass} to generate the equivalent iodine length $L_{io}$ by the forward projection operator $A$. Then, we generate the iodine-induced spectra $S^{(k)}_{io}(E) = S^{(k)}(E)e^{-\mu_{io} L_{io}}$, hence the total X-ray intensity can be modeled as,
\begin{equation}
	\begin{aligned}
		I_t^{(k)} &= \int S^{(k)}_{io}(E)e^{-\mu_{wa} L_{wa}} \text{d}E + I_s
		\label{eq:I_t_io_model}
	\end{aligned}
\end{equation}
where the object becomes purely water-equivalent. Therefore, the scatter can be directly estimated by the scatter estimation method for SMFFS \cite{ZhangPMB2021}. In this paper, we use the residual spectral linearization approach\cite{ZhangPMB2021}.
It should be noted that only data with enough energy separation are used for the scatter estimation. After a regular interpolation, the full-scale scatter distribution is estimated.

\begin{figure*}[htb]
	\centering
	\includegraphics[width=160mm]{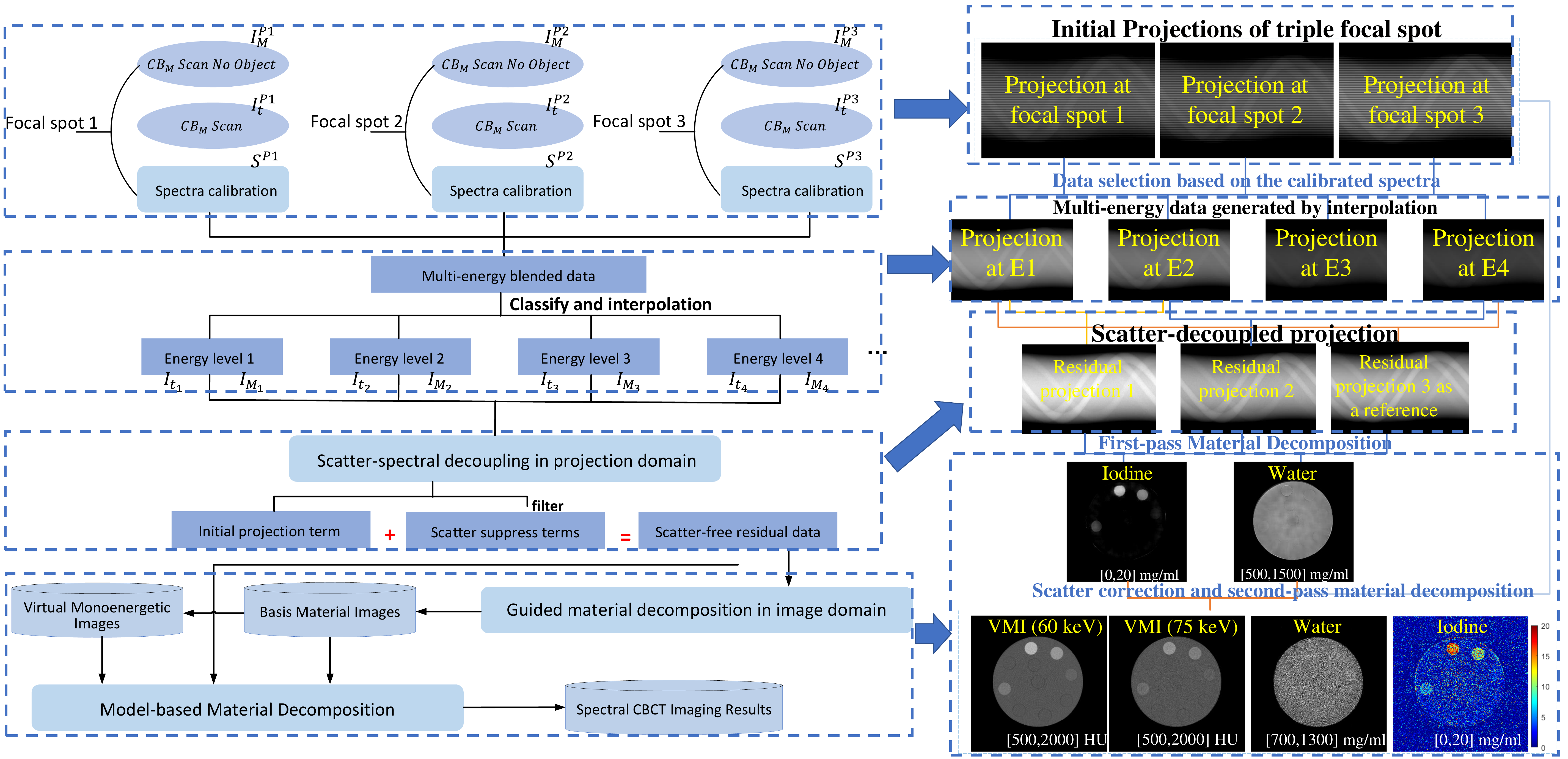}
	\caption{Left:The overall framework of the scatter-decoupled material decomposition method: 1) multi-energy data partitioning and residual projection generation; 2) first-pass material decomposition; 3) scatter estimation; 4) second-pass material decomposition by multi-material spectral correction. Right: Intermediate results} \label{fig:framework}
\end{figure*}

\subsubsection{Second-pass Guided Material Decomposition for Complete Data}
\label{subsubsec:SGMD}
Based on the estimated scatter, we regenerate projection data free from scatter $\hat{P}$ and conduct a second-pass guided material decomposition as follows.

\textbf{1) Spectral imaging for non-penumbra data}:
Basis material images can be generated by a preliminary material decomposition using the non-penumbra projection data free from scatter with enough energy separation $\hat{P}_{L,np}$, $\hat{P}_{H,np}$. Here, for simplicity, we assume a polynomial function is used for the material decomposition as, 
\begin{equation}
	\begin{aligned}
		P_{m1} &= \sum_{i=0}^{N} \sum_{j=0}^{N-i} a_{ij} \cdot \hat{P}_{L,np}^i \cdot \hat{P}_{H,np}^j ,\: a_{00} = 0.\\
		P_{m2} &= \sum_{i=0}^{N} \sum_{j=0}^{N-i} b_{ij} \cdot \hat{P}_{L,np}^i \cdot \hat{P}_{H,np}^j ,\: b_{00} = 0.  
		\label{eq:VMP_np}
	\end{aligned}
\end{equation}
where, $P_{m1}, P_{m2}$ are the projections of the basis materials; the polynomial coefficients $a_{ij}, b_{ij}$ can be generated by polynomial fitting using a series of sample points of basis material densities in advance; the polynomial order $N$ is an empirical parameter (usually 5 is enough). It should be noted that for some large phantom, like an anthropomorphic chest phantom, we can directly use the over-smooth result in Section \ref{subsubsec:firstpass} for $P_{m1},P_{m2}$ to suppress noise and artifacts.

The virtual monochromatic projections (VMP) in non-penumbra area at specific energies $E_L, E_H$ (chosen empirically) can be obtained as, 
\begin{equation}
	\begin{aligned}
		\text{VMP}_{L,np} &= \mu_1 (E_L) \cdot P_{m1} + \mu_2 (E_L) \cdot  P_{m2} \\
		\text{VMP}_{H,np} &= \mu_1 (E_H) \cdot P_{m1} + \mu_2 (E_H) \cdot P_{m2}
		\label{eq:VMP_FP}
	\end{aligned}
\end{equation}

\textbf{2) Spectral imaging for penumbra data}:
Taking advantage of the calibrated spectra in the penumbra area and the preliminary iodine result, $P_{m2}$, the VMP in the penumbra area can also be estimated by using a multi-material spectral calibration (MMSC) method\cite{Gao2014MMC},
\begin{equation}
	\begin{aligned}
		\text{VMP}_{L,p} &= \sum_{i=0}^{N}\sum_{j=0}^{N-i} c_{p,ij} \cdot \hat{P}_p^i \cdot P_{m2}^j , \: c_{00}=0.\\
		\text{VMP}_{H,p} &= \sum_{i=0}^{N}\sum_{j=0}^{N-i} d_{p,ij} \cdot \hat{P}_p^i \cdot P_{m2}^j , \: d_{00}=0.
	\end{aligned}
\end{equation}
where, $c_{p,ij},d_{p,ij}$ can also be generated by the polynomial fitting using a series of sample points of basis material densities in advance; and $P_p$ is the scatter-corrected projection in the penumbra area.

\textbf{3) Spectral reconstruction from the whole data}
To maintain the consistency of the penumbra and non-penumbra data, we blend the VMP in penumbra area $\text{VMP}_{L,np}^{intp}, \text{VMP}_{H,np}^{intp}$ generated via interpolation of $\text{VMP}_{L,np}, \text{VMP}_{H,np}$, and the VMP in penumbra area generated by MMC, $\text{VMP}_{L,p}$, $\text{VMP}_{H,p}$.
For simplicity, a hard threshold is used here,
\begin{equation}
	\text{VMP}_{p,final} = \left\{
	\begin{aligned}
		\text{VMP}_{p}    &, \frac{\lvert \text{VMP}_p-\text{VMP}_{np}^{intp} \rvert}{\text{VMP}_{np}^{intp}}<=t_0 \\
		\text{VMP}_{np}^{intp}   &, \frac{\lvert \text{VMP}_p-\text{VMP}_{np}^{intp} \rvert}{\text{VMP}_{np}^{intp}}>t_0
	\end{aligned}
	\right.
	\label{eq:VMP_blend}
\end{equation}
The threshold $t_0$ is empirically chosen as 0.2 in this paper. The VMPs can be used to  reconstruct VMIs and so do the basis material images after a simple image-domain material decomposition. 

It should be noted that the preliminary material decomposition by the polynomial fitting method may cause some artifacts due to the possibility of slight misalignment of spectral projections in SMFFS. Thus, we also implemented a one-step model-based, material decomposition method (IFBP)\cite{Li2019IFBP} for the second-pass guided material decomposition. In the iterative method, the overall process is similar. In each iteration, we first do material decomposition based on the one-step iterative formula\cite{Li2019IFBP}, and calculate the forward projection $P_{m1}, P_{m2}$; then, we implement the steps from Eq. \eqref{eq:VMP_FP} to \eqref{eq:VMP_blend} as usual; and finally, we update the iteration by the basis material images.

\begin{figure}[htb]
	\centering
	\includegraphics[width=65mm]{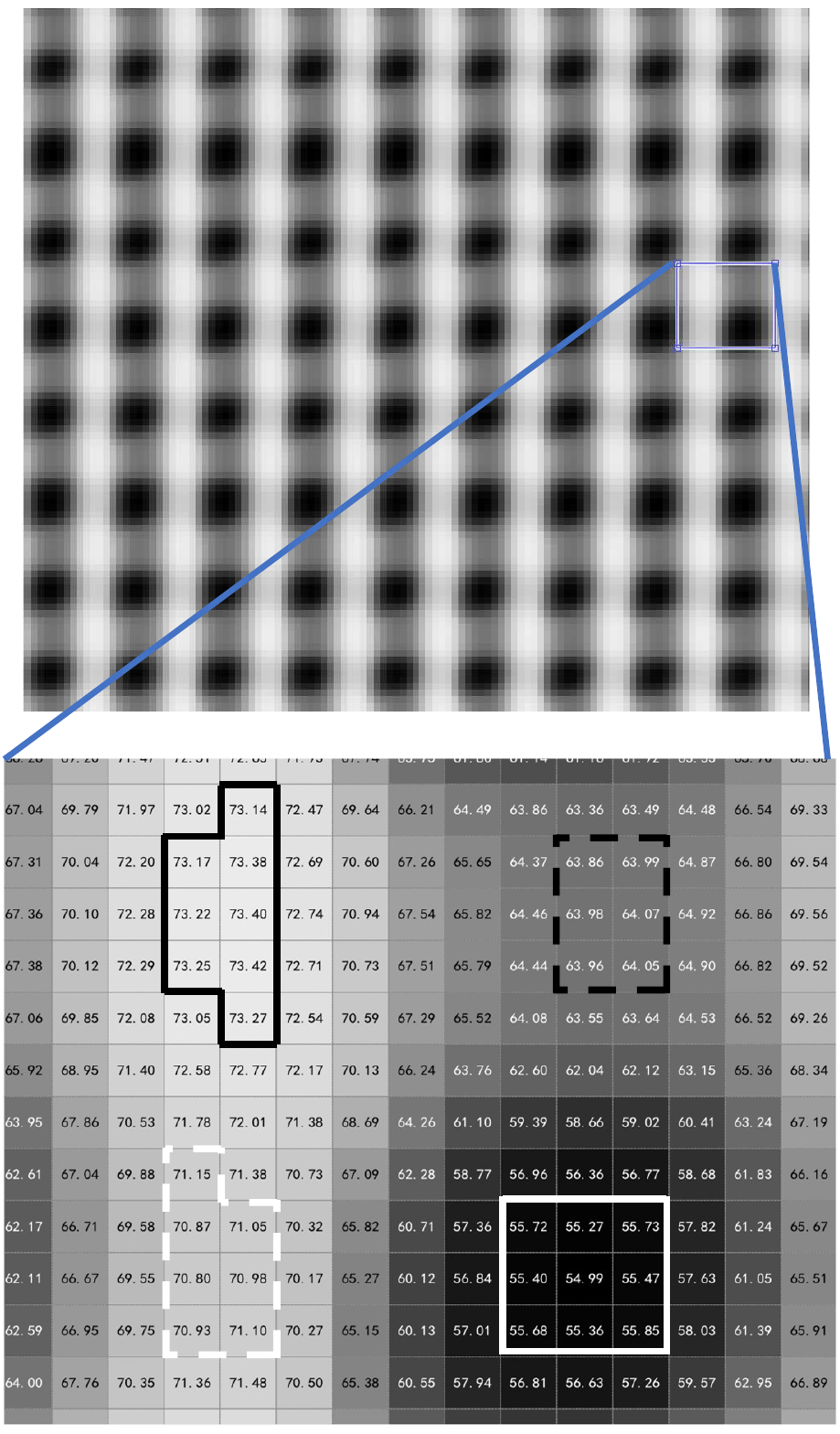}
	\caption{An example of non-penumbra data selection. Top: mean energy of the equivalent spectra, window: [55,75] keV. Bottom: a zoomed-in pattern, white box: original energy (54-56 keV); black dotted box: low energy (63.8-64.1 keV); white dotted box: middle energy (70.5-71.2 keV); black box: high energy (73.1-74 keV).} \label{fig:penumbra_select}
\end{figure}

\subsection{Some Implementation Details}
\label{subsec:algorithm}
Figure \ref{fig:framework} shows the framework of the SDMD method. Here we would like to provide some implementation details. 

\textbf{1) EM-based Spectra Calibration}:
Referring to our previous work\cite{Gao2019spectral}, we first calibrate the spectra without the modulator by an expectation maximization (EM) algorithm\cite{Sidky2006EM}; then we calculate the modulator thicknesses according to the transmission of the modulator under different prefilters; finally, we calculate the spectra with the modulator using the spectra without modulator and the modulator thicknesses.\par

\textbf{2) Adaptive Spectral Data Selection from Triple Focal Spot Positions}:
Based on the mean energy of the calibrated spectra, we mark the data and spectra with subscript $_O,_L,_M,_H$ from triple focal spot positions.
After a rough selection based on the mean energy, some pixels with similar mean energy but with relatively large difference in spectra. The middle energy and the penumbra data may still be blended, as shown in Fig.~\ref{fig:penumbra_select}. Some pixels pointed to by the arrow with the mean energy in the range of middle energy, but obviously belong to the penumbra data. In this case, we use two constraints to segmented out the penumbra data.
First, based on the elaborate geometry design of SMFFS, we can only select the non-penumbra pixels with corresponding non-penumbra pixels under other focal spot positions. Second, we can calculate the spectra error  between the spectra at the low-energy of pixel $i$ with the reference spectrum of low-energy $S_{L}^r$, and $S_{L}^r$ can be predetermined. 
Then, we can set a threshold $T_L$ for the low-energy spectra error, only the pixels with $\frac{\|S_{L}^i-S_{L}^r\|^2_2}{\|S_{L}^r\|^2_2}<T_L$ are selected as the final low-energy non-penumbra data. Similarly, we can also select the non-penumbra data for middle-energy and high-energy data. 
In practice, if no pixel is selected in a period of the modulator, the threshold $T_L$ will be multiplied by a constant, and the constant and $T_L$ can be empirical. Finally, we select and interpolate the data of each category ($S_O,S_L,S_M,S_H$) to obtain the complete multi-energy data. 

\subsection{Validation Cases}
\subsubsection{Setup of Monte Carlo Simulations}
A key assumption of our proposed method is that the X-ray scatter distributions across different focal spot positions possess a strong similarity as long as the distance of focal spot deflection is relatively small. In order to validate that, Monte Carlo simulations by GEANT4 was conducted. 
In Monte Carlo simulations, the detector consisted of $240\times200$ pixels, with a pixel size of 1.5 mm to reduce running time; the phantom was a uniform water cylinder with a diameter of 200 mm and a height of 160 mm; the modulator was a ``Moly. + Moly.'' 2D mixed modulator with the first layer of 0.4 mm thick Mo, and the second layer of 0.2 mm thick Mo. The interval and width of the 1-D strip modulator were set as 0.6 mm and 0.9 mm respectively, which is the same as the manufactured modulator in our physical experiments; the X-ray source was operated at 50 keV. A typical CBCT imaging geometry was simulated, where the source-to-modulator distance (SMD) was 450 mm, the source-to-isocenter distance (SID) was 750 mm, and the source-to-detector distance (SDD) was 1150 mm. %Scattered signals were collected among five different focal spot positions of (0,0,0), (0,-0.7,0), (0,0.7,0), (0,1.4,0), and (0,0,1.4) mm, respectively.

\subsubsection{Setup of Numerical Simulations}
\label{sec:simluation}
Numerical simulations were conducted to validate the feasibility of our method. As a feasibility study, we collected data across three focal spot positions to mimic FFS. The source was operated at 120 kVp in SMFFS scans. A $1440\times 10 -$pixel flat-panel detector was simulated with a pitch of 0.3 mm. The scattered signals in sequential dual-energy scans, and across three focal spot positions in SMFFS were generated by a rough scatter estimation model\cite{Ohnesorge1999scatter} with added fluctuation ($1\pm0.01$), where the SPRs were up to $\sim100$\%. 
Parameters in numerical simulations are summarized in Table~\ref{tab:parameters}. To maintain a consistent dose, the photon flux at the source for one scan of SMFFS was set twice as much as one scan of 80 or 120 kVp due to the attenuation of the modulator. The phantom in numerical simulations was generated by a clinical abdominal CT image data set. For simplicity, we modeled all soft-tissue regions of the abdominal CT image as water-equivalent and added eight iodine contrasts of four different densities (5, 10, 15, and 20 mg/ml iodine) for a quantitative assessment.  

\begin{table}[tb]
	\centering
	\caption[]{\upshape Parameters in numerical simulations and physics experiments.}\label{tab:parameters}
	{
		\begin{tabular}{llll}
			\hline\hline
			\rule{0pt}{8pt}
			&Parameters & Simulation & Experiment\\  
			\hline
			\rule{0pt}{8pt}
			& SDD & 1200 mm & 1200 mm\\
			& SID & 800 mm  & 755 mm\\ 
			& SMD  & 400 mm & 430 mm \\          
			&Views  & 720 & 1206, 750\\
			&FFS distance  & 1.2 mm & 1.32 mm\\
			&\multicolumn{3}{c}{Modulator}\\
			& Material & Moly. & Moly.\\
			&Thickness & \multicolumn{2}{c}{0, 0.2, 0.4, 0.6 mm}\\
			&Spacing &0.6 mm & 0.6 mm\\
			&Period & 1.2 mm & 1.5 mm\\
			
			\hline\hline
		\end{tabular}
	}
\end{table}

\subsubsection{Setup of Physical Experiments}
\begin{figure}[htb]
	\centering
	\includegraphics[width=80mm]{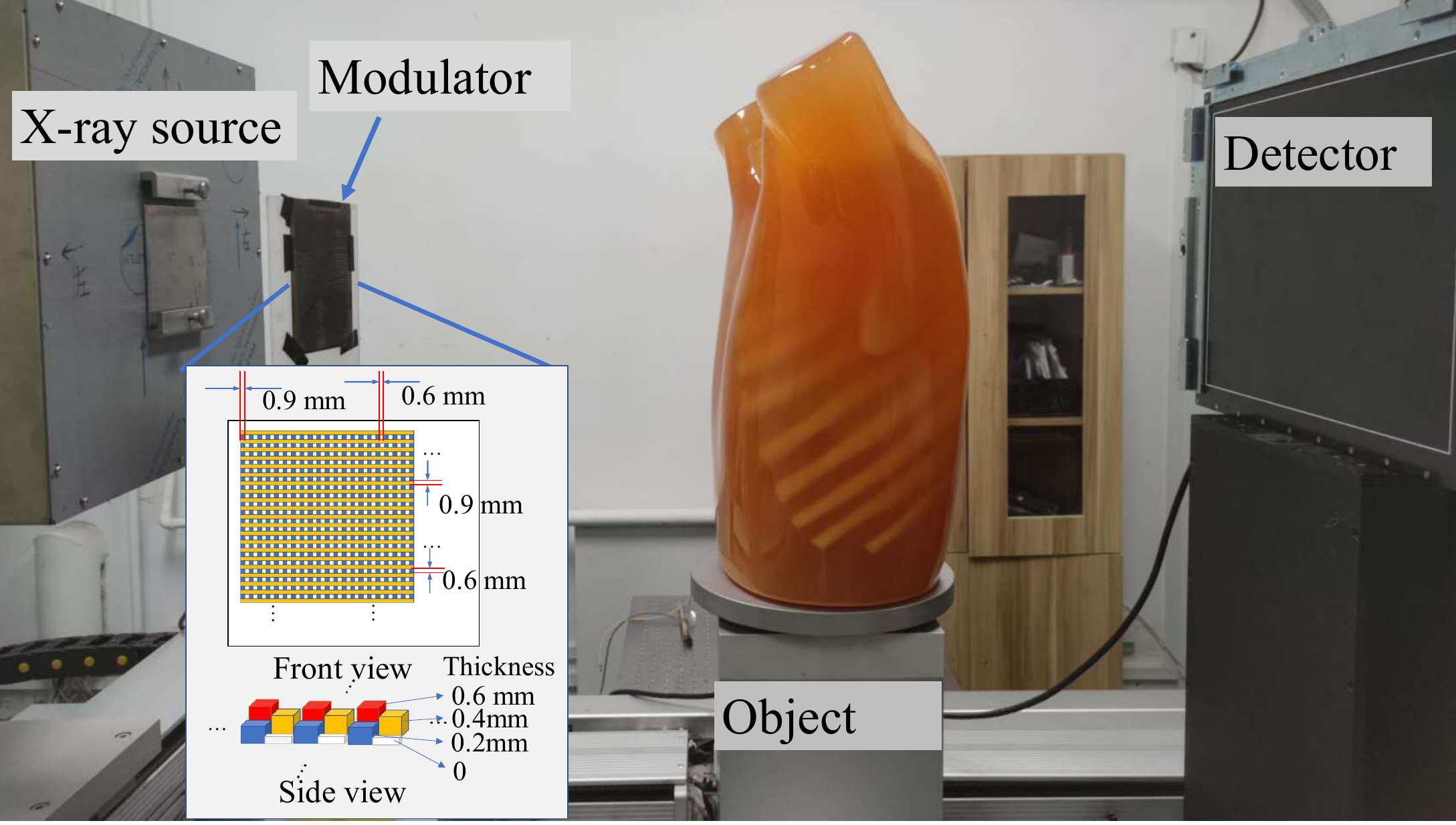}
	\caption{The experimental CBCT platform.} \label{fig:platform}
\end{figure}
Physical experiments were conducted on our tabletop CBCT system as shown in Fig.~\ref{fig:platform}. The X-ray source used a Varex G-242 tube with a focal spot of 0.4 mm; the detector was Varex 4030 DX flat-panel detector;  the mixed 2D modulator was formed by stacking a 0.4 mm of Moly. and a 0.2 mm of Moly. 1D strip modulators, which has four kinds of filters with $0, 0.2, 0.4, 0.6$ mm of Moly. as shown in Fig. \ref{fig:SMFFS_system}, 
and the 1D strip modulators of Moly. both have a spacing of 0.6 mm and a period of 1.5 mm. 
With the X-ray source operated at 120 kVp, the 2-D mixed spectral modulator and in the case of no object, enough energy separation can be achieved by the spectra of the measured data (50/75 keV for lowest and highest mean energy) and residual data (41/59 keV).  The optimization of the modulator design will be discussed in a separate article due to the length of this paper. 

In physical experiments, the source was placed at (0,0,0), (1.32,0,0), and (0,0,1.32) mm sequentially to mimic the flying focal spot deflection; the detector worked in binning 2 mode with $1024 \times 768$ pixels, $ 0.388\times0.388 \ mm^2$ per pixel; a Gammex multi-energy CT phantom was scanned by SMFFS scans (120 kVp, 386 mAs of each scan) and sequential 80/120 kVp dual-energy scans (80 kVp, 760mAs and 120 kVp, 386 mAs) in full-fan scan (the fan gives roughly symmetric coverage on both sides of the rotation axis), and an anthropomorphic Kyoto chest phantom inserted with a 5 mg/ml iodine cylinder was scanned by SMFFS scans (120 kVp, 386 mAs of each scan) and 80/140 kVp with 0.5 mm copper filter dual-energy scans (80 kVp, 800mAs and 140 kVp, 300 mAs) in half-fan scan (the fan has significantly more coverage on one side of the rotation axis than the other, where the detector offset is 13.0 cm).
Table \ref{tab:parameters} summarizes the major parameters.

\subsubsection{Evaluations}
We use the root mean squared error (RMSE) for the CT number in virtual monochromatic image (VMI) as,
\begin{equation}
	\text{RMSE} = \sqrt{\frac{1}{N}\sum_{i=1}^N \frac{\bar{\mu}_{r_i}-\bar{\mu}_{true_i}}{\bar{\mu}_{true_i}}} 
\end{equation}
where $\bar{\mu}_{r_i}$ is the mean value of the ROI $i$, and $\bar{\mu}_{true_i}$ is the theoretical value of the ROI $i$.
Then, in order to evaluate the performance of scatter artifacts reduction, the CT number non-uniformity among selected ROIs within a uniform region of the object is defined as,
\begin{equation}
	\Delta_{\text{MAX}} = \text{max}(\mu_i)-\text{min}(\mu_i)
\end{equation}
where $\text{max}(\mu_i)$ is the maximum value of the average CT numbers in selected ROIs, and $\text{min}(\mu_i)$ is the minimum value.

\section{Results}
\label{sec:Simulations}
\subsection{Monte Carlo Simulations}
\begin{figure}[htb]
	\centering
	\includegraphics[width=60mm]{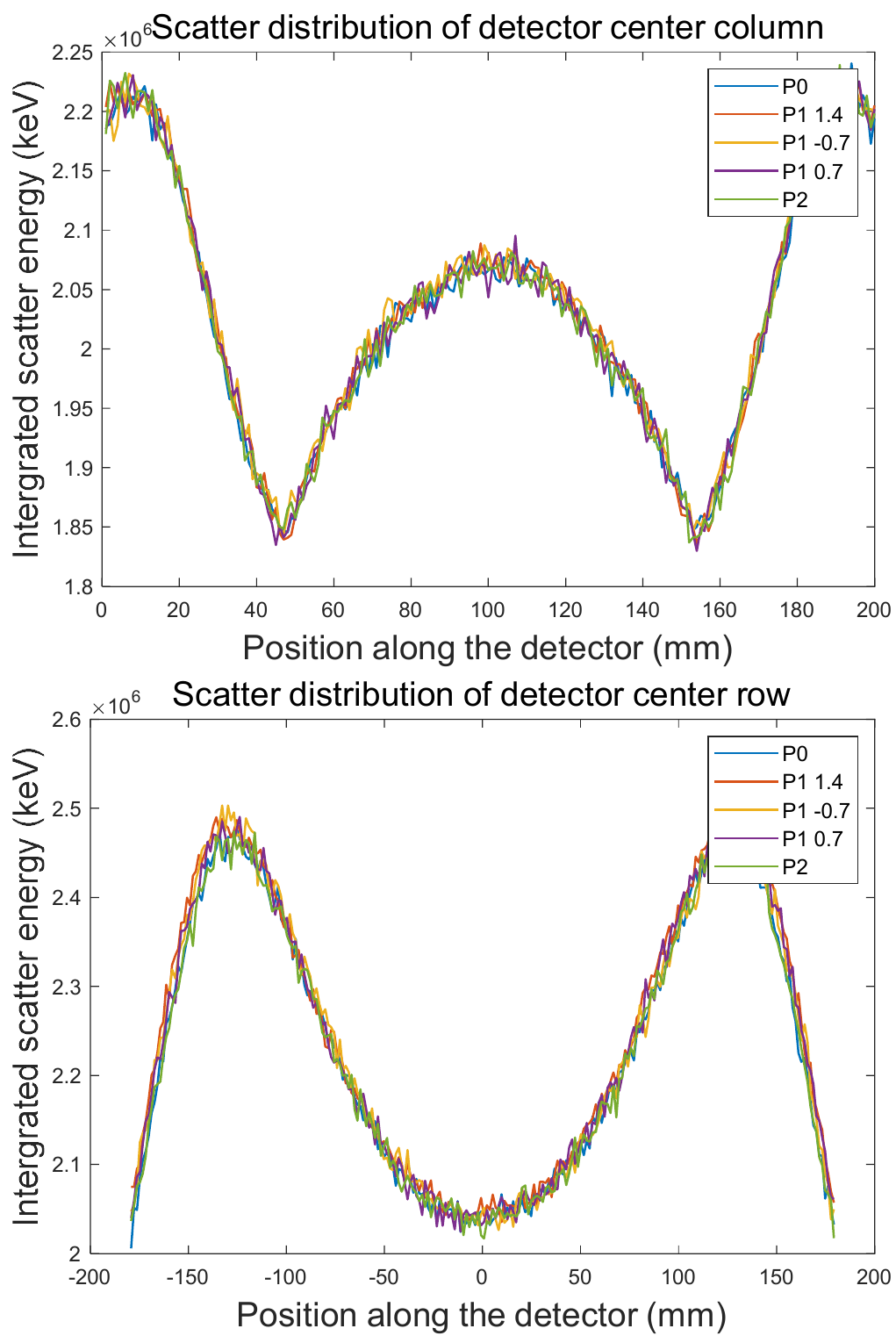}
	\caption{The central horizontal and vertical profiles of the scatter distributions across different focal spot positions in the MC simulation.} \label{fig:MC_results}
\end{figure}
Figure \ref{fig:MC_results} shows the scatter profiles in both vertical and horizontal directions across different focal spot positions in MC simulations, where the focal spot positions include P0 (0,0,0), P1 1.4 (0,1.4,0), P1 -0.7 (0,-0.7,0), P1 0.7 (0,0.7,0), P2 (0,0,1.4).
Using scatter distribution at P0 $I_{s_i}^{P0}$ as a reference, the mean relative errors of all pixels $\frac{\lvert I_{s_i}^{P0} - I_{s_i}^{Pk} \rvert }{I_{s_i}^{P0}}$ are 0.83\%, 0.67\%, 0.66\%, 0.54\% for P1 1.4, P1 -0.7, P1 0.7, P2, respectively, and the maximum relative error is 4.1\%. Strong similarity can be observed in scatter distributions across different focal spot positions with small deflections, validating a key assumption in and an advantage of SMFFS.

\subsection{Numerical Simulations}
\begin{figure*}[htb]
	\centering
	\includegraphics[width=160mm]{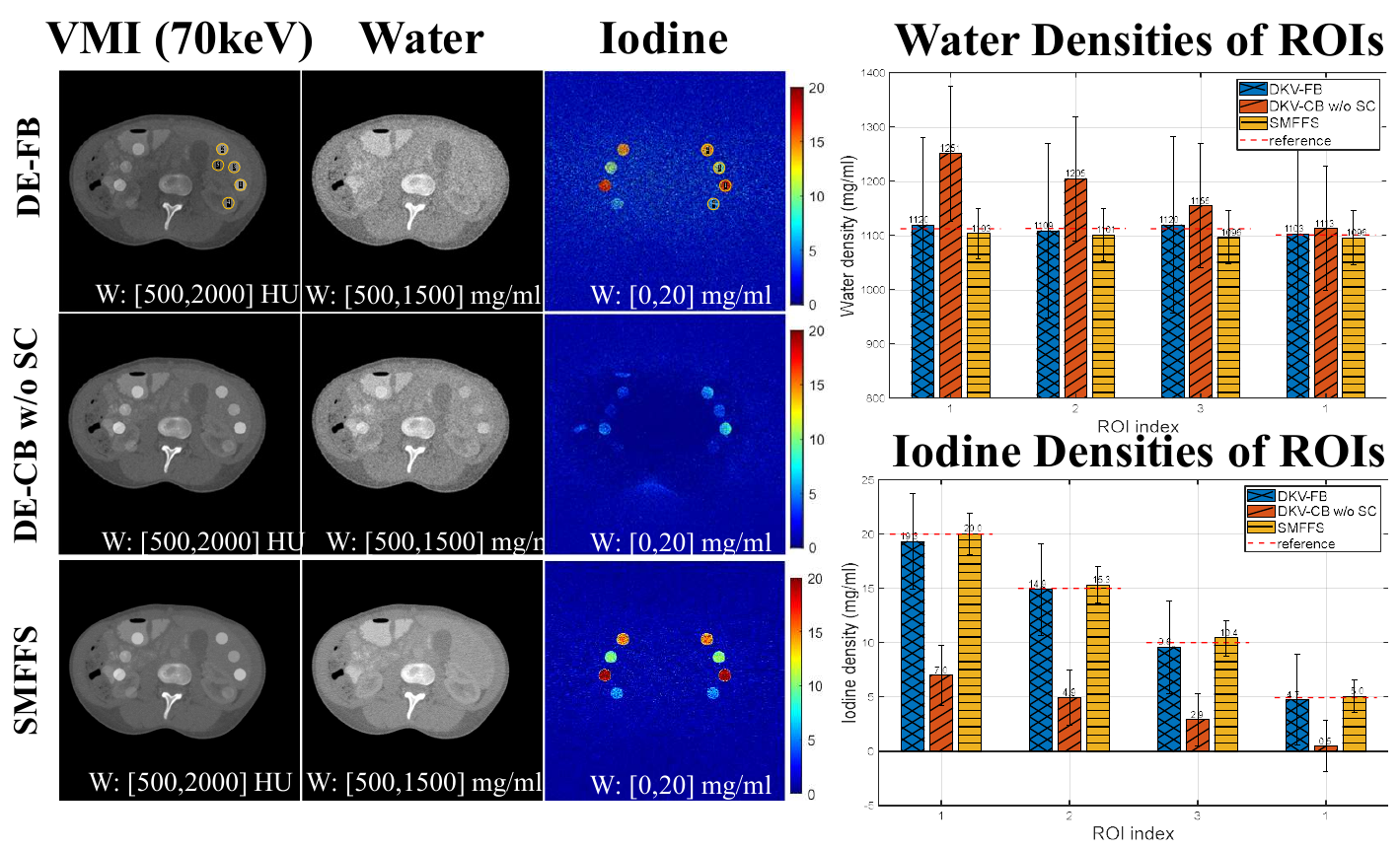}
	\caption{Spectral reconstruction results in simulations. Top row: 80/120 kVp (DKV) scan without scatter; second row: DKV scan with scatter; third row: SMFFS scan with scatter.
	} \label{fig:Ab_MD}
\end{figure*}

Figure \ref{fig:Ab_MD} shows the spectral reconstruction results and quantitative analysis of iodine and water images from SMFFS scans with scatter, and compared with that from the 80/120 kVp dual-energy fan-beam scan (DKV-FB) without scatter and cone-beam scan with scatter but without scatter correction (DKV-CB w/o SC).

%\begin{figure}[htb]
%	\centering
%	\includegraphics[width=70mm]{simul_result_abdomen_ROI_all.pdf}
%	\caption{Spectral reconstruction results in simulations. Top row: 80/120 kVp (DKV) scan without scatter; second row: DKV scan with scatter; third row: SMFFS scan with scatter.
	%		} \label{fig:Ab_MD}
%\end{figure}
\begin{table}[htb]
	\caption{\upshape Averaged CT number (HU) and RMSE for ROIs of VMIs (70keV) in Fig. \ref{fig:Ab_MD}.}
	\centering
	\tabcolsep=3pt
	\begin{tabular}{cccccc}
		\hline\hline
		\rule{0pt}{8pt} 
		ROI  & 1 & 2 & 3 & 4 & RMSE\\
		\hline
		\rule{0pt}{8pt}
		Reference & 1632.2 & 1502.6 & 1372.2 & 1228.9 & 0\\
		DKV-FB  & 1622.3 & 1496.3 & 1369.0 & 1226.2 & 6.3  \\
		DKV-CB & 1433.3  & 1332.5 & 1229.6  & 1112.5 & 157.8  \\
		SMFFS & 1624.1 & 1499.7 & 1367.1 & 1226.7 & 5.1  \\
		\hline\hline
	\end{tabular} \label{tab:evaluation_simul}
\end{table}

Table \ref{tab:evaluation_simul} lists the averaged CT number and RMSEs of ROIs in VMIs at 70 keV. Compared with DKV-CB scans, the RMSE of SMFFS scans is reduced from 157.8 HU to 5.1 HU. These results showed SMFFS using the SDMD method can effectively suppress the scatter artifacts and increase the CT number accuracy.

\subsection{Physical Experiments}
\subsubsection{Gammex Phantom}
\begin{figure*}[htb]
	\centering
	\includegraphics[width=160mm]{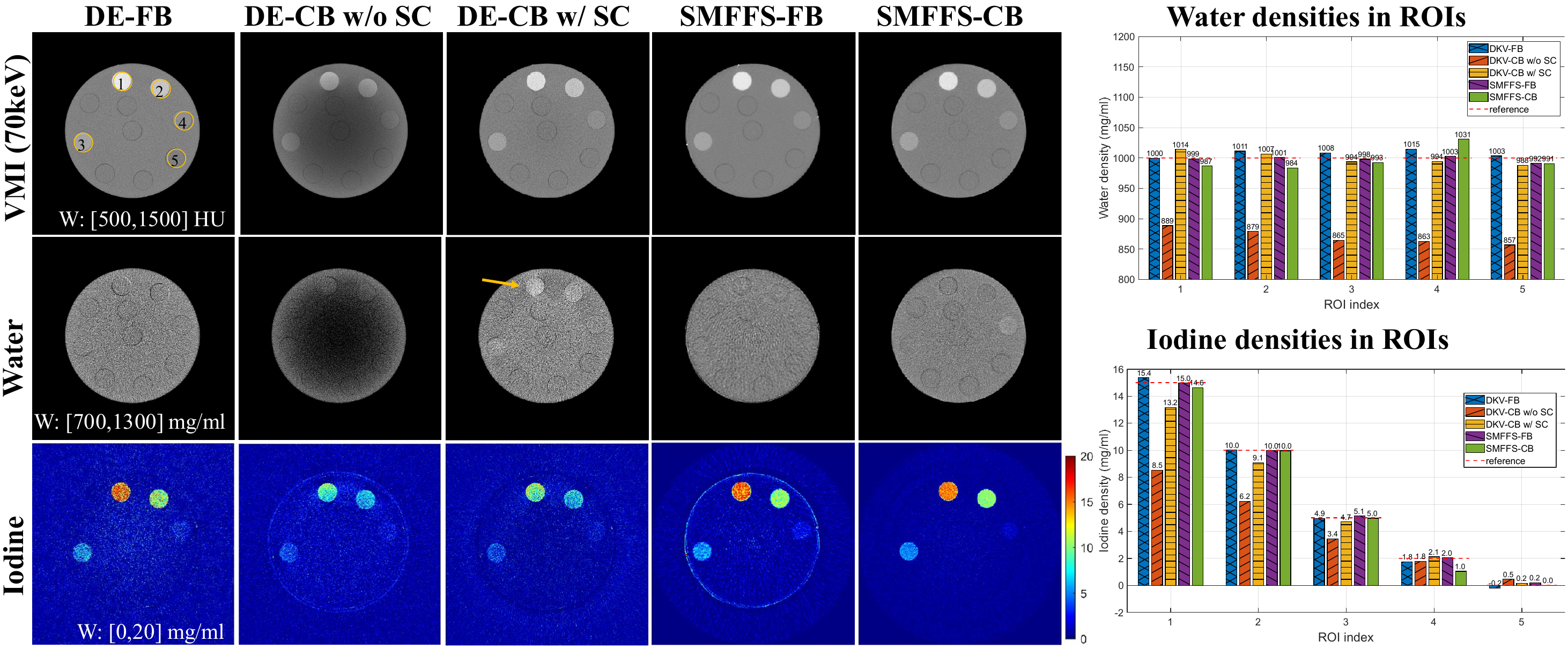}
	\caption{Spectral reconstruction results of the Gammex phantom in experiments. DKV: 80/120 kVp. FB: fan-beam scans; CB w/o SC: cone-beam scans without scatter correction; CB w/ SC: cone-beam scans with scatter correction; SMFFS FB: SMFFS fan-beam scans; SMFFS CB: SMFFS cone-beam scans.} 
	\label{fig:MEphan_experiment}
\end{figure*}

Figure \ref{fig:MEphan_experiment} shows the spectral reconstruction results from SMFFS, and compared with that from sequential dual-energy scans of 80/120 kVp (DKV) with a narrowed collimator (fan beam, FB), a wide collimator (cone beam, CB) without and with scatter correction by a kernel-based method fASKS\cite{Sun2010kernelscatter} available in the CBCT Software Tools (CST) (Varex Imaging Corporation). The references of the quantitative density results in ROIs of iodine and water images are obtained by the user manual of the Gammex Multi-Energy CT Phantom. 

Table \ref{tab:evaluation_exp} shows the averaged CT number and RMSEs of ROI $1-4$ in VMI.
The RMSEs of ROIs in VMI (70 keV) in Fig. \ref{fig:MEphan_experiment} are reduced from 214.3 to 19.2 HU by our method compared with 80/120 kVp dual-energy CBCT without scatter correction. 

\begin{table}[htb]
	\caption{\upshape Averaged CT number (HU) and RMSE for ROIs of VMIs (70keV) in Fig. \ref{fig:MEphan_experiment}.}
	\centering
	\tabcolsep=2pt
	\begin{tabular}{cccccc}
		\hline\hline
		\rule{0pt}{8pt}
		ROI  & 1 & 2 & 3 & 4 & RMSE \\
		\hline
		\rule{0pt}{8pt}
		Reference & 1397.1 & 1263.3 & 1132.7 & 1052.4 & 0 \\
		DKV-FB  & 1499.6 & 1271.6 & 1137.0 & 1060.1 & 6.2 \\
		DKV-CB w/o SC & 1110.5  & 1041.2 & 954.1  & 909.5  & 214.3 \\
		DKV-CB w/ SC & 1356.8 & 1243.0 & 1116.2 & 1049.5 & 24.1 \\
		SMFFS-FB & 1388.0 & 1260.5 & 1131.5 & 1056.1 & 5.2 \\
		SMFFS-CB & 1367.4 & 1242.5 & 1121.8 & 1058.4 & 19.2 \\
		\hline\hline
	\end{tabular} \label{tab:evaluation_exp}
\end{table}

\subsubsection{Kyoto Chest Phantom}
\begin{figure*}[htb]
	\centering
	\includegraphics[width=160mm]{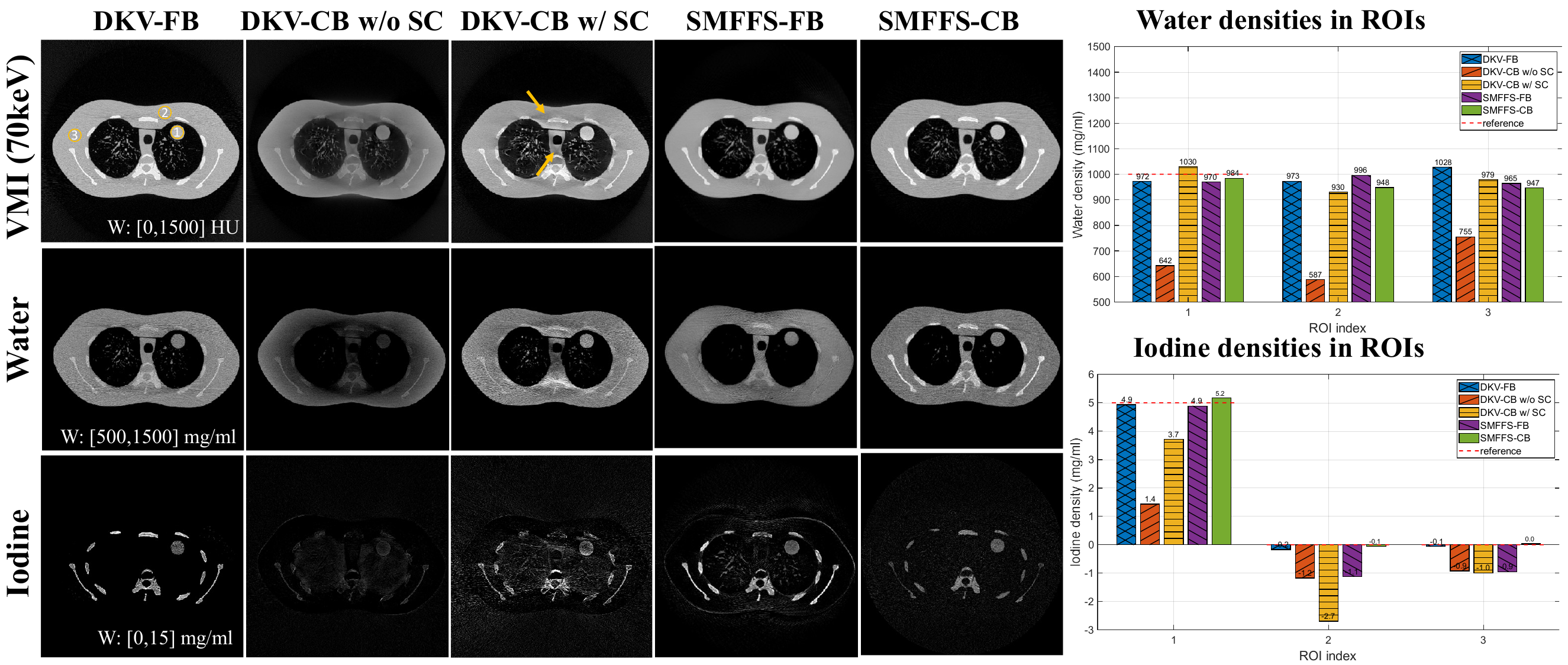}
	\caption{Spectral reconstruction results of the chest phantom in experiments. DKV-FB: 80 / 140 kVp fan-beam scans; DKV-CB w/o SC: 80 / 140 kVp cone-beam scans without scatter correction; DKV-CB w SC: 80 / 140 kVp cone-beam scans with scatter correction; SMFFS FB: SMFFS fan-beam scans; SMFFS CB: SMFFS cone-beam scans.} 
	\label{fig:chest_experiment}
\end{figure*}

Figure \ref{fig:chest_experiment} shows the performance comparison of the anthropomorphic chest phantom by SMFFS scans and 80 / 140 kVp dual energy (DKV) scans. The estimated material densities for 5 mg/ml iodine are 5.01 for DKV-FB, 1.59 for DKV-CB, 3.71 for DKV-CB with scatter correction, 4.96 for SMFFS-FB, 5.23 for SMFFS-CB, respectively. And the area pointed by the yellow arrow in DKV-CB with traditional scatter correction shows the artifacts of the material decomposition in the chest anatomies, indicating the sensitivity of material decomposition to the scatter correction, which can be suppressed by our method. The right column of Fig. \ref{fig:chest_experiment} also shows the water and iodine densities of the selected ROIs (marked by the yellow circle).

\begin{figure*}[htb]
	\centering
	\includegraphics[width=160mm]{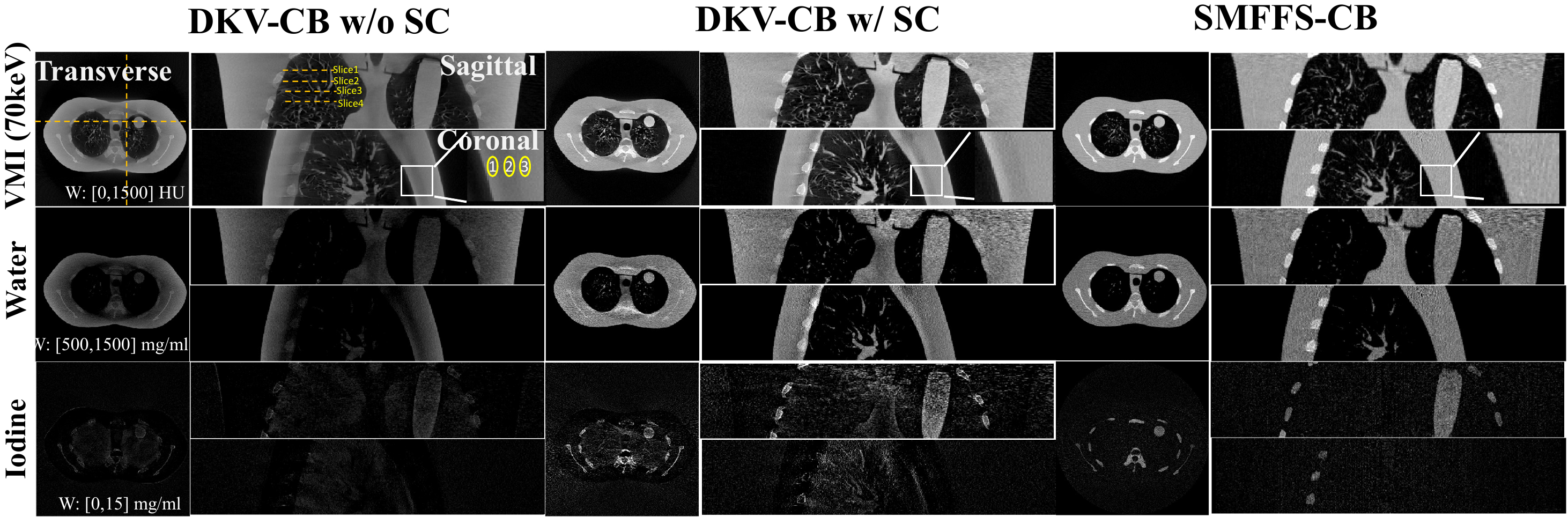}
	\caption{VMI (70 keV), iodine, and water images of the chest phantom in sagittal and coronal view by the sequential dual-energy and SMFFS cone-beam scans. Display window: iodine: [0,15] mg/ml, water: [500,1500] mg/ml, VMI(70keV): [0,1500] HU.}
	\label{fig:chest_experiment_3D}
\end{figure*}

Figure \ref{fig:chest_experiment_3D} shows the sagittal and coronal view of the cone-beam imaging results, where the yellow line in the sagittal image is the ROI for 5 mg/ml iodine cylinder in different slices. Table \ref{tab:evaluation_3D_chest} shows the averaged CT number and RMSEs of the 5 mg/ml iodine cylinder of several slices in VMIs.
And Table \ref{tab:chest_3D_uniformity} shows the averaged CT number and non-uniformity of ROIs in VMIs at the coronal view.
It is seen that a great quantitative imaging performance and consistent performance of SMFFS can be achieved, with the scatter artifacts effectively suppressed, and the non-uniformity can be reduced from 184.0 to 14.1 HU. 

\begin{table}[h]
	\caption{\upshape Averaged CT number (HU) and RMSE for 5 mg/ml iodine cylinder of VMIs (70keV) at different z-slices as shown in Fig. \ref{fig:chest_experiment_3D}.}
	\centering
	\tabcolsep=3pt
	\begin{tabular}{cccccc}
		\hline\hline
		\rule{0pt}{8pt}
		Slices  & 1 & 2 & 3 & 4 & RMSE\\
		\hline
		\rule{0pt}{8pt}
		Reference & 1130.0 & 1130.0 & 1130.0 & 1130.0 & 0\\
		DKV-CB w/o SC & 667.3  & 695.0 & 696.4 & 713.2 & 437.6
		\\
		DKV-CB w/ SC & 1123.0 & 1134.9 & 1121.2 & 1103.7 & 14.5
		\\
		SMFFS-CB & 1119.8 & 1118.3 & 1115.6& 1118.6 & 11.8
		\\
		\hline\hline
	\end{tabular} \label{tab:evaluation_3D_chest}
\end{table}

\begin{table}[h]
	\caption{\upshape Averaged CT number (HU) and non-uniformity for ROIs in Fig. \ref{fig:chest_experiment_3D}.}
	\centering
	\tabcolsep=3pt
	\begin{tabular}{cccc}
		\hline\hline
		\rule{0pt}{8pt}
		ROI  & DKV-CB w/o SC & DKV-CB w/ SC & SMFFS-CB \\
		\hline
		\rule{0pt}{8pt}
		1 & 589.4 & 862.5 & 945.4
		\\
		2 & 658.1 & 878.7 & 951.9
		\\
		3 & 773.4 & 921.9 & 959.5
		\\
		$\Delta_{\text{MAX}}$ & 184.0 & 59.4 & 14.1
		\\
		\hline\hline
	\end{tabular} \label{tab:chest_3D_uniformity}
\end{table}

As we can see, these preliminary spectral reconstruction results and quantitative analysis above demonstrate that our proposed SDMD method can significantly improve the image quality and the quantitative performance of CBCT, even better than that of the sequential dual-energy cone-beam results in our preliminary cases.

\section{Discussions and Conclusions} 
%\vspace{-25em}
\label{sec:Conclusions}
It is highly desired for CBCT to achieve better quantitative performance, even though its spectral imaging capability is still limited by challenges to obtain an accurate scatter correction and material decomposition from dual- or multi-energy spectral CBCT data. In this paper, we proposed a scatter-decoupled material decomposition (SDMD) method for spectral CBCT imaging with SMFFS. As a preliminary study, numerical simulations and physical experiments showed that the scatter-spectral intertwined problem in spectral CBCT can be simultaneously solved by our method.

There are some limitations of our current study, in particular regarding the penumbra effect, noise, and SMFFS data acquisition.

a) Penumbra effect: a major factor that limits the spectral CBCT imaging performance in SMFFS, which could lead to relatively sparse spectral data and inaccurate estimation of the system spectra. In this work, to overcome the  sparse spectral problem in the penumbra region, we adopted the multi-material spectral correction method in Section \ref{subsubsec:SGMD} in order to maximize the usage of collected data. A ring fix approach\cite{Wang2023AAPM} was also applied to remove ring artifact caused by residual errors in spectra estimation mainly occurring in the penumbra regions. Theoretically, a precise model for the focal spot and the spectral modulator may achieve better results. However, extended focal spot size and shape, detector glare, manufacturing defects of the modulator, and other non-ideal factors may occur in reality, making it complex to accurately model the whole system. A smaller focal spot size would suppress the penumbra effect, but may lead to higher system cost and lower X-ray tube power. A more precise spectrum estimation method for the penumbra area will be helpful.
% it's not a fair comparison for an iterative method for SMFFS and a polyfitting for DKV.
b) Dose and noise: the dose for DKV scans and SMFFS scans are similar because the modulator is placed between the source and the object. Regard the dose of a scan without the modualtor and with the setup in SMFFS scans as 1, the total dose of SMFFS scans (3-FFS) is about 1.5, while the total dose of DKV scans is about 1.7 for the Gammex phantom and 1.9 for the chest phantom. In this paper, we did not compare the noise between SMFFS scans and DKV scans, because it's not a fair comparison for SMFFS scans using an iterative method and DKV scans only using a polynomial fitting method, in which case the noise performance of SMFFS scans are obviously better than DKV scans. And it should be noted that the noise in the SDMD method is mainly increased by the scatter-decoupled step and the material decomposition due to the relatively moderate energy separation by SMFFS. In this paper, we separated the scatter-decoupled term and low-pass filtered the scatter suppressing term, which is able to reduce noise with spatial resolution preserved. We also used a similarity-constraint to suppress the noise in the first-pass material decomposition. 

c) Data acquisition: In this work, we moved the focal spot sequentially to mimic the flying focal spot view-by-view, and this may not take full advantage of the scatter properties. In practice, one can take just a portion of views with triple focal spot deflections (3-FFS) to get the best estimate of scatter, and then interpolate the scatter distribution to the other views. As a result, most of views can be operated only with dual focal spot deflections (2-FFS). By using such a scan strategy, data collection can be more efficient. 

Also, it should be noted that in this work, the scatter correction and spectral imaging performance for dual-kVp CBCT scan may be further improved if a more sophisticated scatter correction method can be applied instead of the relatively straightforward kernel-based one used in our experiments. In this preliminary study, for simplicity the dual-kVp dual-energy scans of the Gammex phantom and the chest phantom were operated at 80/120 kVp and 80/140 kVp, respectively. In general, the best selection of the low- and high-kVp combination for a specific application requires significant efforts of optimizations in system design and algorithm development.

Future work includes further optimization of the design of the modulator and the SDMD algorithm, developing deep-learning based material decomposition for SMFFS, with more phantom evaluations and analyses. Also, it would be very interesting to explore the possibility of integrating the SMFFS method into other existing spectral CT technologies such as the dual-layer detector and fast-kV switching.

\section*{Acknowledgments}
This project was supported in part by grants from the National Natural Science Foundation of China (No. 12075130 and No. U20A20169), and in part by the National Key R\&D Program of China (No. 2022YFE0131100)

% following only if there is an appendix

\section*{References}
\addcontentsline{toc}{section}{\numberline{}References}
\vspace*{-20mm}

% Following assumes you are using bibtex. However, for submission to the
% journal you MUST explicitly INCLUDE THE REFERENCES IN THE TEX FILE. 
% In that case you need the following

% \begin{thebibliography}{10}
% insert the .bbl file generated by bibtex here
	%This will be a series of entries from your .bib file formatted
	%something like
	%\bibitem{Me09}
        %{I.~Meijsing, B.~W.~Raaymakers, A.~J.~E.~Raaijmakers \it et al.},
        %\newblock {Dosimetry for the MRI accelerator: the impact of a 
	%magnetic field on the response of a Farmer NE2571 ionization chamber},
        %\newblock Phys. Med. Biol. {\bf 54}, 2993 -- 3002 (2009).

% \end{thebibliography}

% The following is when using bibtex and picks up the example.bib file

%\bibliography{Explicit address of .bib file}
\bibliography{./Ref}      %example.bib is on the same directory

\begin{thebibliography}{10}

\bibitem{Posiewnik2019IGRT}
M.~Posiewnik and T.~Piotrowski,
\newblock A review of cone-beam CT applications for adaptive radiotherapy of
  prostate cancer,
\newblock Physica Medica-European Journal of Medical Physics {\bf 59}, 13--21
  (2019).

\bibitem{Dental2021review}
T.~Kaasalainen, M.~Ekholm, T.~Siiskonen, and M.~Kortesniemi,
\newblock Dental cone beam CT: An updated review,
\newblock Physica Medica-European Journal of Medical Physics {\bf 88}, 193--217
  (2021).

\bibitem{Connell2021CBCTBreast}
A.~M. O'Connell, T.~J. Marini, and D.~T. Kawakyu-O'Connor,
\newblock Cone-Beam Breast Computed Tomography: Time for a New Paradigm in
  Breast Imaging,
\newblock Journal of Clinical Medicine {\bf 10} (2021).

\bibitem{McCollough2020multi}
C.~H. McCollough, K.~Boedeker, D.~Cody, X.~Duan, T.~Flohr, S.~S. Halliburton,
  J.~Hsieh, R.~R. Layman, and N.~J. Pelc,
\newblock Principles and applications of multienergy CT: Report of AAPM Task
  Group 291,
\newblock Med Phys {\bf 47}, e881--e912 (2020).

\bibitem{Rajendran2022FirstClinicalPCD}
K.~Rajendran, M.~Petersilka, A.~Henning, E.~R. Shanblatt, B.~Schmidt, T.~G.
  Flohr, A.~Ferrero, F.~Baffour, F.~E. Diehn, L.~Yu, P.~Rajiah, J.~G. Fletcher,
  S.~Leng, and C.~H. McCollough,
\newblock First Clinical Photon-counting Detector CT System: Technical
  Evaluation,
\newblock Radiology {\bf 303}, 130--138 (2022).

\bibitem{Muller2016kVCBCT}
K.~Muller, S.~Datta, M.~Ahmad, J.~H. Choi, T.~Moore, L.~Pung, C.~Niebler, G.~E.
  Gold, A.~Maier, and R.~Fahrig,
\newblock Interventional dual-energy imaging-Feasibility of rapid kV-switching
  on a C-arm CT system,
\newblock Med Phys {\bf 43}, 5537 (2016).

\bibitem{Cassetta2020kVCBCT}
R.~Cassetta, M.~Lehmann, M.~Haytmyradov, R.~Patel, A.~Wang, L.~Cortesi,
  D.~Morf, D.~Seghers, M.~Surucu, H.~Mostafavi, and J.~C. Roeske,
\newblock Fast-switching dual energy cone beam computed tomography using the
  on-board imager of a commercial linear accelerator,
\newblock Phys Med Biol {\bf 65}, 015013 (2020).

\bibitem{ShiDLCBCT2020}
L.~X. Shi, M.~H. Lu, N.~R. Bennett, E.~Shapiro, J.~Zhang, R.~Colbeth,
  J.~Star-Lack, and A.~S. Wang,
\newblock Characterization and potential applications of a dual-layer
  flat-panel detector,
\newblock Med Phys {\bf 47}, 3332--3343 (2020).

\bibitem{Stahl2021DLCBCT}
F.~Stahl, D.~Schafer, A.~Omar, P.~van~de Haar, F.~van Nijnatten, P.~Withagen,
  A.~Thran, E.~Hummel, B.~Menser, A.~Holmberg, M.~Soderman, A.~Falk~Delgado,
  and G.~Poludniowski,
\newblock Performance characterization of a prototype dual-layer cone-beam
  computed tomography system,
\newblock Med Phys {\bf 48}, 6740--6754 (2021).

\bibitem{Ji2021PCDCBCT}
X.~Ji, M.~Feng, K.~Treb, R.~Zhang, S.~Schafer, and K.~Li,
\newblock Development of an Integrated C-Arm Interventional Imaging System With
  a Strip Photon Counting Detector and a Flat Panel Detector,
\newblock IEEE Trans Med Imaging {\bf 40}, 3674--3685 (2021).

\bibitem{Euler2016Inv}
A.~Euler, A.~Parakh, A.~L. Falkowski, S.~Manneck, D.~Dashti, B.~Krauss,
  Z.~Szucs-Farkas, and S.~T. Schindera,
\newblock Initial Results of a Single-Source Dual-Energy Computed Tomography
  Technique Using a Split-Filter: Assessment of Image Quality, Radiation Dose,
  and Accuracy of Dual-Energy Applications in an In Vitro and In Vivo Study,
\newblock Investigative radiology {\bf 51}, 491--498 (2016).

\bibitem{Petrongolo2018}
M.~Petrongolo and L.~Zhu,
\newblock Single-Scan Dual-Energy CT Using Primary Modulation,
\newblock IEEE Trans Med Imaging {\bf 37}, 1799--1808 (2018).

\bibitem{Tivnan2019SPIE}
M.~Tivnan, S.~Tilley~Ii, and J.~W. Stayman,
\newblock Physical Modeling and Performance of Spatial-Spectral Filters for CT
  Material Decomposition,
\newblock Proc SPIE Int Soc Opt Eng {\bf 10948} (2019).

\bibitem{Gao2019SMFFS}
H.~Gao, C.~Wu, and S.~Wang,
\newblock Spectral modulator with flying focal spot: a new concept of
  full-scale multi-energy cone-beam CT and simultaneous scatter correction,
\newblock in {\em AAPM 61st Annual Meeting}, 2019.

\bibitem{Hsieh2021FFS}
S.~Hsieh, N.~Bennett, and A.~Wang,
\newblock Dual Energy CT Using Flying Focal Spot and An Anti-Scatter Grid as a
  Filter,
\newblock in {\em AAPM 63rd Annual Meeting}, 2021.

\bibitem{Stayman2021MP}
J.~W. Stayman, M.~Tivnan, W.~Wang, N.~Shapira, G.~J. Gang, and P.~B. Noël,
\newblock Spectral CT using a fine grid structure and varying x-ray incidence
  angle,
\newblock Med Phys {\bf 48}, 6412--6420 (2021).

\bibitem{zhu2006scatter}
L.~Zhu, N.~R. Bennett, and R.~Fahrig,
\newblock Scatter correction method for X-ray CT using primary modulation:
  theory and preliminary results,
\newblock IEEE Trans Med Imaging {\bf 25}, 1573--87 (2006).

\bibitem{Gao2010scatter}
H.~Gao, R.~Fahrig, N.~R. Bennett, M.~Sun, J.~Star-Lack, and L.~Zhu,
\newblock Scatter correction method for x-ray CT using primary modulation:
  phantom studies,
\newblock Med Phys {\bf 37}, 934--46 (2010).

\bibitem{Ritschl2015}
L.~Ritschl, R.~Fahrig, M.~Knaup, J.~Maier, and M.~Kachelrieß,
\newblock Robust primary modulation-based scatter estimation for cone-beam CT,
\newblock Med Phys {\bf 42}, 469--78 (2015).

\bibitem{Chen2016modu}
Y.~Chen, Y.~Song, J.~Ma, and ZhaoJ.,
\newblock Optimization-based scatter estimation using primary modulation for
  computed tomography,
\newblock Med Phys {\bf 43}, 4753 (2016).

\bibitem{PivotTMI2020}
O.~Pivot, C.~Fournier, J.~Tabary, J.~M. Letang, and S.~Rit,
\newblock Scatter Correction for Spectral CT Using a Primary Modulator Mask,
\newblock IEEE Trans Med Imaging {\bf 39}, 2267--2276 (2020).

\bibitem{shi2022singleshot}
L.~Shi, N.~R. Bennett, and A.~S. Wang,
\newblock Single-Shot Quantitative X-ray Imaging Using a Primary Modulator and
  Dual-Layer Detector: Simulation and Phantom Studies,
\newblock Proceedings of SPIE--the International Society for Optical
  Engineering {\bf 12031}, 1203106 (2022).

\bibitem{Deng2020CTMeeting}
Y.~Deng and H.~Gao,
\newblock Triple-energy x-ray CT using spectral modulator with flying focal
  spot: modulator design and scatter-modeled material decomposition,
\newblock in {\em Proceedings of the 6th Int. Conference on Image Formation in
  X-Ray CT}, pages 66--69, 2020.

\bibitem{Deng2023fully3d}
Y.~Deng and H.~Gao,
\newblock Design of Scatter-Decoupled Material Decomposition for Multi-Energy
  Blended CBCT Using Spectral Modulator with Flying Focal Spot,
\newblock in {\em Proceedings of the 17th Virtual International Meeting on
  Fully 3D Image Reconstruction in Radiology and Nuclear Medicine}, pages
  78--81, 2023.

\bibitem{KachelrieTMI2006}
M.~Kachelriess, M.~Knaup, C.~Penssel, and W.~A. Kalender,
\newblock Flying focal spot (FFS) in cone-beam CT,
\newblock IEEE Transactions on Nuclear Science {\bf 53}, 1238--1247 (2006).

\bibitem{TivnanMP2021}
M.~Tivnan, W.~Wang, and J.~W. Stayman,
\newblock A prototype spatial–spectral CT system for material decomposition
  with energy‐integrating detectors,
\newblock Medical Physics {\bf 48}, 6401--6411 (2021).

\bibitem{GaoSPIE2020}
H.~Gao, H.~Zhou, L.~Zhu, N.~R. Bennett, and A.~S. Wang,
\newblock {\em Spectral modulator with flying focal spot for cone-beam CT: a
  feasibility study}, volume 11312 of {\em SPIE Medical Imaging},
\newblock SPIE, 2020.

\bibitem{GaoMP2021}
H.~Gao, T.~Zhang, N.~R. Bennett, and A.~S. Wang,
\newblock Densely sampled spectral modulation for x-ray CT using a stationary
  modulator with flying focal spot: a conceptual and feasibility study of
  scatter and spectral correction,
\newblock Med Phys {\bf 48}, 1557--1570 (2021).

\bibitem{Wang2016PMB}
T.~Wang and L.~Zhu,
\newblock Dual energy CT with one full scan and a second sparse-view scan using
  structure preserving iterative reconstruction (SPIR),
\newblock Phys Med Biol {\bf 61}, 6684--6706 (2016).

\bibitem{ZhangPMB2021}
T.~Zhang, Z.~Chen, H.~Zhou, N.~R. Bennett, A.~S. Wang, and H.~Gao,
\newblock An analysis of scatter characteristics in x-ray CT spectral
  correction,
\newblock Phys Med Biol {\bf 66}, 075003 (2021).

\bibitem{Gao2014MMC}
H.~Gao, A.~Cohen, and Y.~Imai,
\newblock Quantitative uniformity of iodinated contrast across the z-coverage
  of large cone-angle CT,
\newblock in {\em the 3rd Int. Conference on Image Formation in X-Ray CT},
  pages 220--223, 2014.

\bibitem{Li2019IFBP}
M.~Li, Y.~Zhao, and P.~Zhang,
\newblock Accurate Iterative FBP Reconstruction Method for Material
  Decomposition of Dual Energy CT,
\newblock IEEE Trans Med Imaging {\bf 38}, 802--812 (2019).

\bibitem{Gao2019spectral}
H.~Gao, L.~Zhang, R.~Grimmer, and R.~Fahrig,
\newblock Physics-based spectral compensation algorithm for x-ray CT with
  primary modulator,
\newblock Phys Med Biol {\bf 64}, 125006 (2019).

\bibitem{Sidky2006EM}
H.~Gao, L.~Zhang, R.~Grimmer, and R.~Fahrig,
\newblock A robust method of x-ray source spectrum estimation from transmission
  measurements: Demonstrated on computer simulated, scatter-free transmission
  data,
\newblock Journal of Applied Physics {\bf 97}, 124701--124701 (2005).

\bibitem{Ohnesorge1999scatter}
B.~Ohnesorge, T.~Flohr, and K.~Klingenbeck-Regn,
\newblock Efficient object scatter correction algorithm for third and fourth
  generation CT scanners,
\newblock Eur Radiol {\bf 9}, 563--9 (1999).

\bibitem{Sun2010kernelscatter}
M.~Sun and J.~M. Star-Lack,
\newblock Improved scatter correction using adaptive scatter kernel
  superposition,
\newblock Phys Med Biol {\bf 55}, 6695--720 (2010).

\bibitem{Wang2023AAPM}
Y.~Wang, H.~Gao, Z.~Chen, and L.~Zhang,
\newblock A Fast and Robust Ring Artifact Removal Method Based on Gradient
  Domain Optimization,
\newblock in {\em AAPM 65th Annual Meeting}, 2023.

\end{thebibliography}
% above points to where we find the master reference list
% and also causes the bibliography to be printed

% When creating your bibliography you should run bibtex on your local
% computer after running pdflatex on your .tex file. bibtex will
% generate a .bbl file.
% Copy the contents of this .bbl file into your main latex document,
% replacing the "\bibliography" command which was pointing at your .bib file.

% following defines style of .bbl file 

%\bibliographystyle{explicit relative path to medphy.bst}
\bibliographystyle{./medphy.bst}    %if this is installed on your system,
				    %it is not essential to have the    ./

% Note that you need to typeset once, then run bibtex, then typeset another
% two times to get the references working properly.

\end{document}